\documentclass[preprint,superscriptaddress,amsmath,amssymb,amsart,aps,prl]{revtex4-1}

\usepackage{graphicx}
\usepackage{dcolumn}
\usepackage{bm}
\usepackage{hyperref}
\usepackage{xcolor}
\usepackage{tikz}
\usepackage{comment}
\usepackage{xr}
\usepackage{cleveref}
\usepackage{multibib}
\usepackage{amsmath}
\newcites{SM}{SM References}
\Crefname{figure}{Fig.}{Figs.}
\Crefname{equation}{Eq.}{Eqs.}
\hypersetup{
    pdfborder={0 0 0}
}

\definecolor{lime}{HTML}{A6CE39}
\DeclareRobustCommand{\orcidicon}{
	\begin{tikzpicture}
	\draw[lime, fill=lime] (0,0) 
	circle [radius=0.16] 
	node[white] {{\fontfamily{qag}\selectfont \tiny ID}};
	\draw[white, fill=white] (-0.0625,0.095) 
	circle [radius=0.007];
	\end{tikzpicture}
	\hspace{-2mm}
}
\foreach \x in {A, ..., Z}{\expandafter\xdef\csname orcid\x\endcsname{\noexpand\href{https://orcid.org/\csname orcidauthor\x\endcsname}
			{\noexpand\orcidicon}}
}

\newcolumntype{C}[1]{>{\centering\arraybackslash}p{#1}}

\usepackage{lineno}
\newcommand*{\WM }{The College of William \& Mary, Williamsburg, Virginia 23185, USA}

\newcommand*{\ANL }{Argonne National Laboratory, Lemont, Illinois 60439, USA}

\newcommand*{\MSU }{Mississippi State University, Mississippi State, Mississippi 39762, USA}

\newcommand*{\TEMP }{Temple University, Philadelphia, Pennsylvania 19122, USA}

\newcommand*{\JLAB }{Thomas Jefferson National Accelerator Facility, Newport News, Virginia 23606, USA}

\newcommand*{\BOULDER }{University of Colorado Boulder, Boulder, Colorado 80309, USA}

\newcommand*{\YER }{A.I. Alikhanyan  National  Science  Laboratory \\ (Yerevan  Physics
Institute),  Yerevan  0036,  Armenia}

\newcommand*{\CUA }{Catholic University of America, Washington, DC 20064, USA}

\newcommand*{\REG }{University of Regina, Regina, Saskatchewan S4S 0A2, Canada}

\newcommand*{\ZAG }{University of Zagreb, Zagreb, Croatia}

\newcommand*{\HU }{Hampton University, Hampton, Virginia 23669, USA}

\newcommand*{\CNU }{Christopher Newport University, Newport News, Virginia 23606, USA}

\newcommand*{\UVA }{University of Virginia, Charlottesville, Virginia 22903, USA}

\newcommand*{\NCAT }{North Carolina A \& T State University, Greensboro, North Carolina 27411, USA}

\newcommand*{\SUNO }{Southern University at New Orleans, New Orleans, Louisiana 70126, USA}

\newcommand*{\UTENN }{University of Tennessee, Knoxville, Tennessee 37996, USA}

\newcommand*{\UCONN }{University of Connecticut, Storrs, Connecticut 06269, USA}

\newcommand*{\ODU }{Old Dominion University, Norfolk, Virginia 23529, USA}

\newcommand*{\OHIO }{Ohio University, Athens, Ohio 45701, USA}

\newcommand*{\UOY }{University of York, Heslington, York, YO10 5DD, UK}

\newcommand*{\FIU }{Florida International University, University Park, Florida 33199, USA}

\newcommand*{\JMU }{James Madison University, Harrisonburg, Virginia 22807, USA}

\newcommand*{\VM}{Virginia Military Institute, Lexington, Virginia 24450, USA}

\newcommand*{\SBU }{Stony Brook University, Stony Brook, New York 11794, USA}
\newcommand*{\TU}{Tsinghua University, Beijing 100084, China}

\begin{document}


\title{Charged kaon and proton multiplicities in semi-inclusive deep-inelastic scattering with 11 GeV electrons}




\author{P.\,Bosted}\affiliation{\WM}
\author{W.\,Armstrong}\affiliation{\ANL} 
\author{H.\,Bhatt}\affiliation{\MSU}
\author{D.\,Dutta}\affiliation{\MSU}
\author{R.\,Ent}\affiliation{\JLAB}   
\author{D.\,Gaskell}\affiliation{\JLAB} 
\author{S.\,Jia}\affiliation{\TEMP}
\author{E.\,Kinney}\affiliation{\BOULDER} 
\author{H.\,Mkrtchyan}\affiliation{\YER}

\author{S.\,Ali}\affiliation{\CUA}
\author{R.\,Ambrose}\affiliation{\REG} 
\author{D.\,Androic}\affiliation{\ZAG}  
\author{C.\,Ayerbe Gayoso}\affiliation{\MSU} 
\author{A.\,Bandari}\affiliation{\WM}   
\author{V.\,Berdnikov}\affiliation{\CUA}    
\author{D.\,Bhetuwal}\affiliation{\MSU}
\author{D.\,Biswas}\affiliation{\HU}  
\author{M.\, Boer}\affiliation{\TEMP}
\author{E.\,Brash}\affiliation{\CNU}    
\author{A.\,Camsonne}\affiliation{\JLAB}
\author{M.\,Cardona}\affiliation{\TEMP}          
\author{J.\,P.\,Chen}\affiliation{\JLAB}           
\author{J.\,Chen}\affiliation{\WM}    
\author{M.\,Chen}\affiliation{\UVA}             
\author{E.\,M.\,Christy}\affiliation{\HU}          
\author{S.\,Covrig}\affiliation{\JLAB}           
\author{S.\,Danagoulian}\affiliation{\NCAT}     
\author{M.\,Diefenthaler}\affiliation{\JLAB}     
\author{B.\,Duran}\affiliation{\TEMP}            
\author{M.\,Elaasar}\affiliation{\SUNO}
\author{C.\,Elliot}\affiliation{\UTENN}
\author{H.\,Fenker}\affiliation{\JLAB}           
\author{E.\,Fuchey}\affiliation{\UCONN}           
\author{J.\,O.\,Hansen}\affiliation{\JLAB}           
\author{F.\,Hauenstein}\affiliation{\ODU}       
\author{T.\,Horn}\affiliation{\CUA}             
\author{G.\,M.\,Huber}\affiliation{\REG}       
\author{M.\,K.\,Jones}\affiliation{\JLAB}          
\author{M.\,L.\,Kabir}\affiliation{\MSU}
\author{A.\,Karki}\affiliation{\MSU}            
\author{B.\,Karki}\affiliation{\OHIO} 
\author{S.\,J.\,D.\,Kay}\affiliation{\REG}\affiliation{\UOY}
\author{C.\,Keppel}\affiliation{\JLAB}           
\author{V.\,Kumar}\affiliation{\REG}
\author{N.\,Lashley-Colthirst}\affiliation{\HU}        
\author{W.\,B.\,Li}\affiliation{\WM}               
\author{D.\,Mack}\affiliation{\JLAB}              
\author{S.\,Malace}\affiliation{\JLAB}           
\author{P.\,Markowitz}\affiliation{\FIU}       
\author{M.\,McCaughan}\affiliation{\JLAB}
\author{E.\,McClellan}\affiliation{\JLAB}
\author{D.\,Meekins}\affiliation{\JLAB}          
\author{R.\,Michaels}\affiliation{\JLAB}         
\author{A.\,Mkrtchyan}\affiliation{\YER}        
\author{G.\,Niculescu}\affiliation{\JMU}        
\author{I.\,Niculescu}\affiliation{\JMU}        
\author{B.\,Pandey}\affiliation{\HU}\affiliation{\VM}           
\author{S.\,Park}\affiliation{\SBU}             
\author{E.\,Pooser}\affiliation{\JLAB}           
\author{B.\,Sawatzky}\affiliation{\JLAB}          
\author{G.\,R.\,Smith}\affiliation{\JLAB}             
\author{H.\,Szumila-Vance}\affiliation{\JLAB}\affiliation{\FIU}
\author{A.\,S.\,Tadepalli}\affiliation{\JLAB}
\author{V.\,Tadevosyan}\affiliation{\YER}        
\author{R.\,Trotta}\affiliation{\CUA}           
\author{H.\,Voskanyan}\affiliation{\YER}
\author{S.\,A.\,Wood}\affiliation{\JLAB}            
\author{Z.\, Ye}\affiliation{\ANL}\affiliation{\TU}
\author{C.\,Yero} \affiliation{\FIU}  
\author{X.\,Zheng}\affiliation{\UVA}        
\collaboration{for the Hall C SIDIS Collaboration}
\noaffiliation

\date{\today}

\begin{abstract}

Measurements of SIDIS  multiplicities for charged kaons and protons from proton and deuteron targets are reported on a grid of hadron kinematic variables $0.3<z<0.7$ and $P_{t}<0.6$ GeV for leptonic variables $0.3<x<0.6$ and $3<Q^2<6$ GeV$^2$. Data were acquired in 2018-2019 at Jefferson Lab Hall C with 10.2 and 10.6~GeV electron beams impinging on 10-cm-long liquid hydrogen and deuterium targets. Electrons (hadrons) were detected in the HMS (SHMS) spectrometers. Multiplicities were fitted for  each bin in $(x,~Q^2,~z,~P_{t})$ as $M_0[1+A\cos(\phi^*)+B\cos(2\phi^*)]$. The kaon kinematic range spans the regions where transverse-momentum-dependent factorization can be applied in SIDIS, and a `soft' central region where other processes are of critical importance. The kaon to pion ratios of $M_0$ are in reasonable agreement with predictions using the DSS fragmentation functions for $K^+$, but are mostly well  below them for $K^-$. The kaon azimuthal modulations are consistent with zero. The kinematic range for protons is centered on the `soft' central region. The proton-to-pion multiplicity ratios are more than an order-of-magnitude larger than TMD predictions at the lowest value of $W^2$, decreasing to as little as a factor of two at the highest value of $W^2$. No significant difference is observed between proton and deuteron targets. These trends are consistent with Lund Monte Carlo predictions. The proton  values of $A$ are consistently positive, with an average value of approximately 0.01, while $B$ is consistent with zero.

\end{abstract}

      
\maketitle

\section{Introduction}
The process of semi-inclusive deep-inelastic electroproduction of 
hadrons (SIDIS) has a long history 
in the study of nucleon structure (see Ref.~\cite{Bacchetta17} 
for a recent review). As discussed in the pioneering paper 
of Berger~\cite{Berger:1987zu}, under suitable kinematic
conditions, the process can be described as
a convolution of quark parton distribution functions 
$F(x,Q^2)$ and fragmentation 
functions $D(z,Q^2)$, both of which follow QCD 
evolution due to gluon radiation, where $(x,Q^2)$ 
are the usual deep-inelastic-scattering (DIS) 
variables, and $z$ is the ratio of the hadron momentum $P_h$ to the 
virtual photon energy $\nu$.
In recent decades parton distribution functions have been developed to include the dependence on the quark transverse momentum $k_t$ as well as the 
longitudinal momentum fraction $x$,  
and fragmentation functions that depend on the hadron longitudinal momentum fraction $z$ as well as the hadron transverse 
momentum $P_{hT}$.
The  SIDIS process therefore offers the potential to
probe the three-dimensional (3D) internal structure of the nucleon, particularly for 
detected pions. The affinity to transverse-momentum-dependent (TMD)
studies is highest for detected
pions, less so for detected kaons, and very small for detected protons 
at Jefferson Lab kinematics~\cite{Boglione:2016bph,Boglione:2019nwk,Boglione:2022gpv}.

In this paper, we follow up on a recent publication of SIDIS pion 
production with 10.6 GeV electrons at Jefferson Lab \cite{HallCSIDIS:2025plr} to present 
results for kaon and proton production in the deep-inelastic scattering
kinematic range of four-momentum transfer squared $3<Q^2<6$ GeV$^2$, 
parton momentum fraction $0.3<x<0.6$, and final-state invariant mass squared
$4<W^2<10.9$ GeV$^2$. Kaon SIDIS
has been reported previously from HERMES with 27 GeV leptons 
(Refs.~\cite{HERMES:2012uyd,HERMES:2012kpt}) at somewhat higher values of $W^2$.
Kaon SIDIS results from COMPASS~\cite{COMPASS:2016crr} 
used 160 GeV muons and focused on the sea-quark region ($x<0.3$) and
much higher values of $W^2$.
We are not aware of any existing  
publications reporting cross sections or multiplicities of identified protons in SIDIS kinematics. 

After a discussion of the formalism and relevance of the present experiment to TMD studies,
we briefly discuss the experimental setup, followed by results for the
kaon and proton.

\section{Formalism and Relevance to TMD studies}

The ``Berger criteria" for factorization to be an appropriate
description of SIDIS can be summarized as $W^2 \gtrsim 10$ GeV$^2$, where $W^2$ is the 
invariant mass squared of the electron-target system defined by $W^2=m^2 + Q^2(1-x)/x$, 
where $m$ is the nucleon mass, for $z>0.2$. Berger also discussed the criteria that
$\Delta y_h\equiv y_h - y_f>2$, where the standard definition of hadron rapidity is 
$y_h= \frac{1}{2}
\ln{(E_h+P_h^L)/(E_h-P_h^L)}$ (longitudinal is defined by the direction of the virtual photon $\bf\vec q$) and the final-state quark rapidity is
given by $y_f = -\ln (\sqrt{Q^2}/m)$ in the Breit frame.

More recently, Boglione {\it et al.}~\cite{Boglione:2016bph,Boglione:2019nwk,Boglione:2022gpv}
have investigated the region in which transverse-momentum-dependent (TMD) parton and fragmentation function factorization can be used as a framework to analyze SIDIS data. They divided SIDIS into three regions: a current fragmentation region ($y_h<-1$) in which factorization should be valid; a target fragmentation region in which the detected hadron
originates from fragmentation of the spectator quarks, described by extended
fracture functions ($y_h>1$); and a soft central region
in between these two, for which no factorization theorems have yet been determined.
They propose the use of a modified hadron rapidity variable given by~\cite{Boglione:2016bph}
\[
y_h =
\ln \left[
\frac{Q z \left( Q^2 - x_n^2 m^2 \right)}
{2 x_n^2 m^2 M_{hT}}
-
\frac{Q}{x_n m}
\sqrt{
\frac{
z^2 \left( Q^2 - x_n^2 m^2 \right)^2
}
{
4 x_n^2 m^2 M_{hT}^2
}
-1
}
\right]
\]
\noindent where 
$$x_n=\frac{2x}{ 1 + \sqrt{1 + 4 x^2 m^2/ Q^2}}$$
is the Nachtmann variable, 
$M_{ht}=\sqrt{P_t^2+M_h^2}$, $M_h$ is the detected
hadron mass, and $P_t$ is the hadron transverse 
momentum with respect to the momentum transfer 
four-vector $q$, and the $y_i = \ln (\sqrt{Q^2}/m)$ 
is initial quark rapidity. 

\begin{figure*}[htb!]
\includegraphics[width=0.9\textwidth]{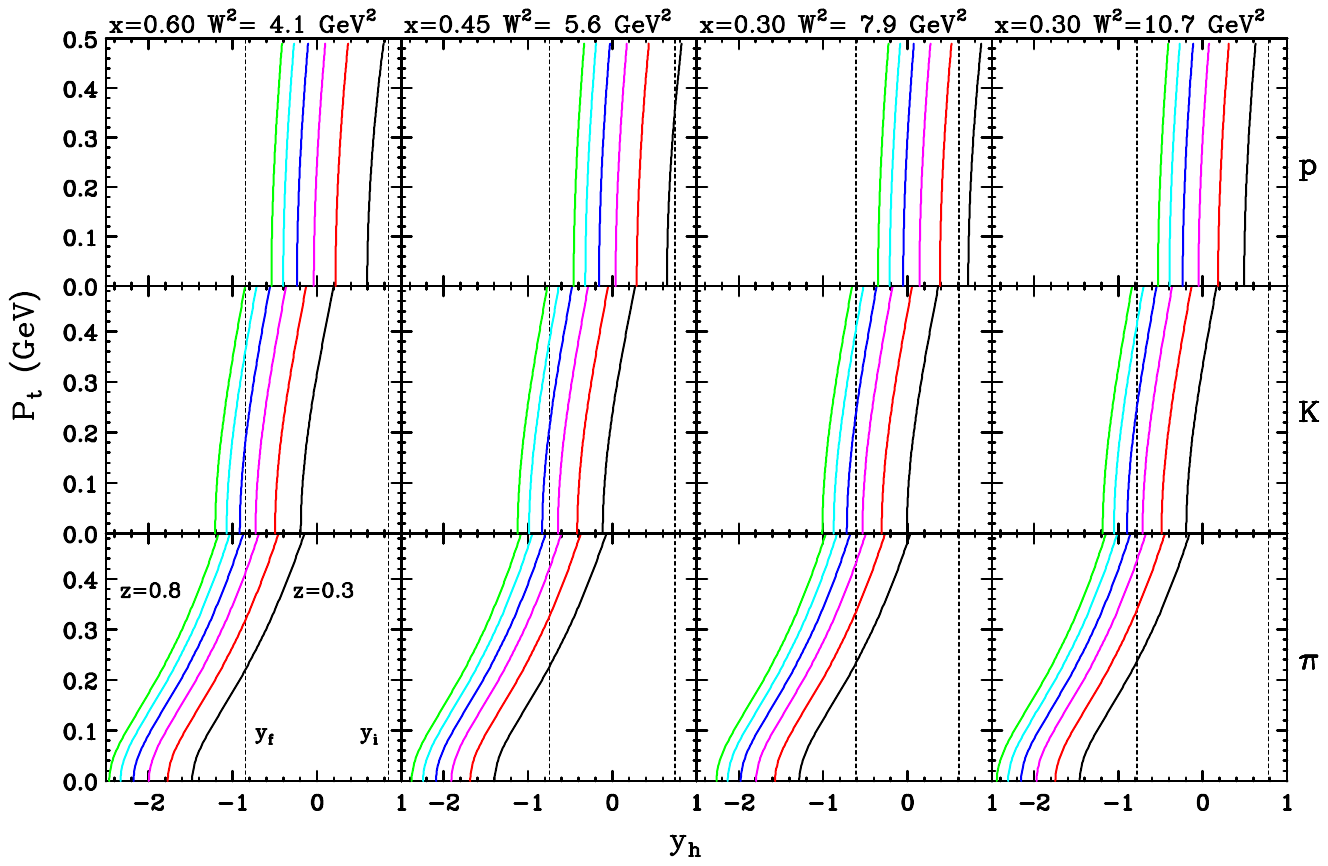} 
    \caption{The relationship between hadron rapidity $y_h$ and hadron transverse momentum $P_t$ for $z$  = 0.3, 0.4, 0.5, 0.6, 0.7, and 0.8 (colored lines, from right to left), for four values of $x,Q^2,W^2$ (columns) and three detected particle flavors (rows). The left and right vertical dashed lines indicate the final-state quark rapidity $y_f$ and the initial-state quark rapidity $y_i$, respectively.
 \label{fig:rap}}
\end{figure*}

To examine the validity of  TMD factorization in the
Boglione {\it et al.} framework, within the kinematic range of the present experiment,  we plot 
in Fig.~\ref{fig:rap} the variables
$y_i$, $y_f$, and $y_h$ as a function of $P_t$ for five values of $z$ at the four kinematic settings of the experiment. For detected pions, there is a substantial range
in $(z,P_t$) for which $y_h<-1$ (although 
$y_f$ is typically about -0.8), and the affinity
to TMD factorization is more than 80\%. However,
for protons, $-1<y_h<1$ for essentially
the entire $z$ and $P_t$ range of the present experiment, implying 
that the data lie in the
soft central region, for which no clear interpretation is available. For the intermediate
case of the kaon, $y_h<-1$ for the larger values
of $z$, as long as $P_t<0.3$ GeV. 

\section{The Experiment}
\subsection{Overview}
\begin{figure}[hbt!]
\centering
\includegraphics[width=1.0\textwidth] {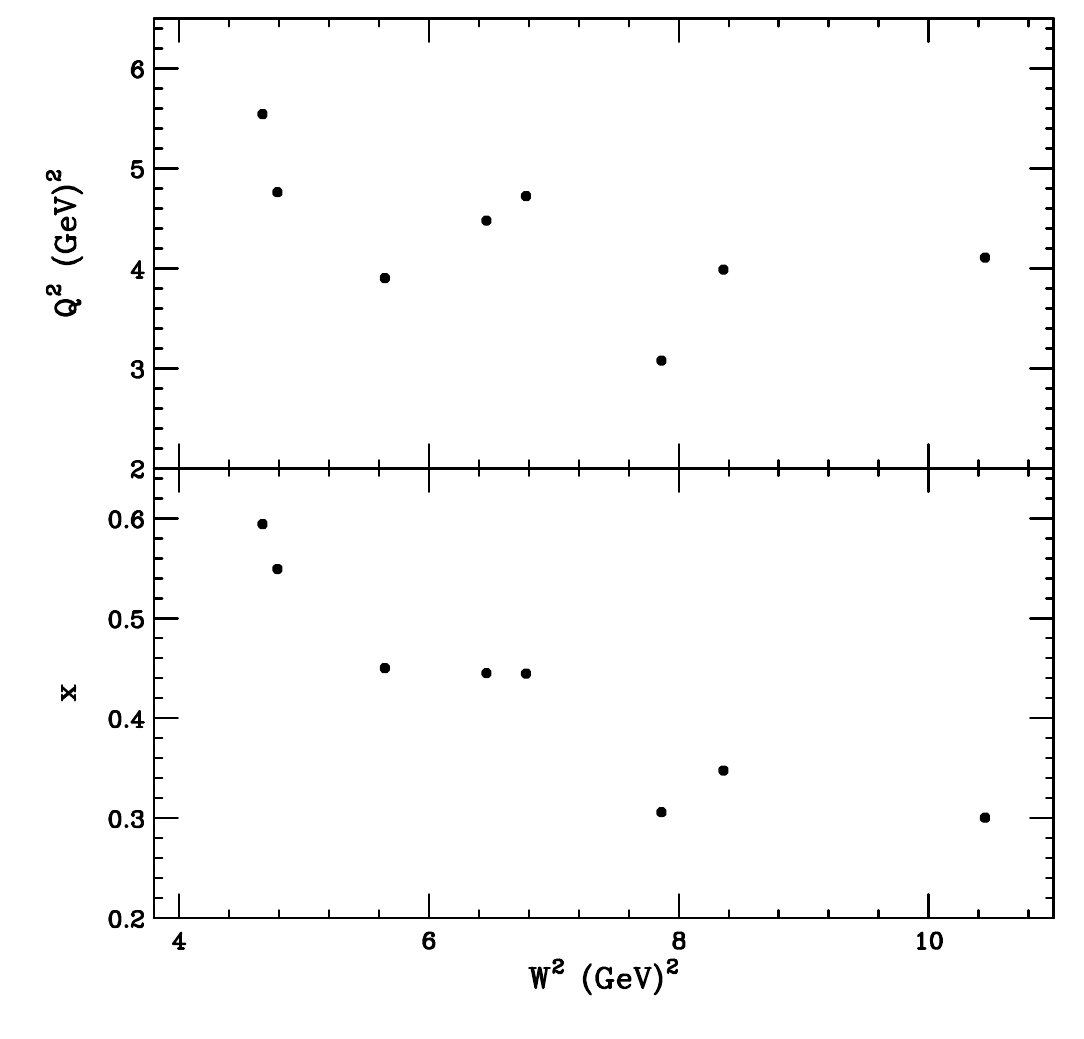}
 \caption{Values of $x$ and $Q^2$ as a function of $W^2$
 for the eight kinematic settings in this paper.}
 \label{fig:kin}
 \end{figure}

The experiment was carried out in 2018-2019, in 
Hall C at Jefferson Lab (JLab) using the quasi-continuous wave electron beam with beam 
energies of 10.2 and 10.6~GeV. The description of the electron beam, target, spectrometers, and detectors is 
given in Refs.~\cite{HallCSIDIS:2025plr,FF_paper}. The kaon and proton results
presented in this paper are for the same eight kinematic settings as in
Ref.~\cite{FF_paper}, and the 
range of $x$ and $Q^2$ 
are plotted as a function of final-state invariant 
mass squared, $W^2$, in Fig.~\ref{fig:kin}. The only things that differed from the pion 
analysis~\cite{HallCSIDIS:2025plr,FF_paper}
are in the particle identification (PID) 
criteria used to identify detected
kaons and protons, as well as the cross section models used in the
evaluation of radiative corrections.
 
\subsection{Kaon identification and simulation}
\begin{figure}[hbt!]
\centering
\includegraphics[width=1.0\textwidth] {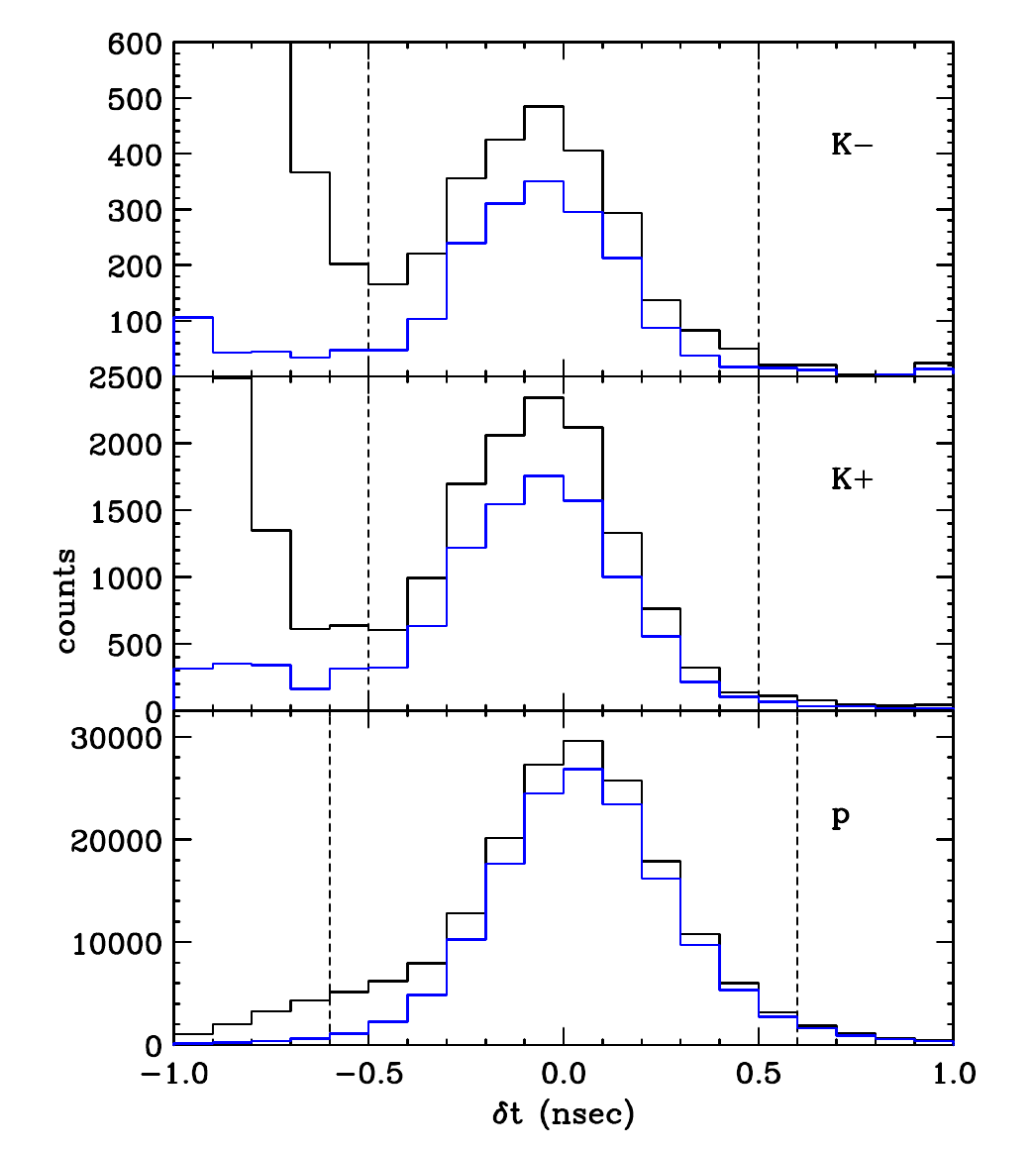}
 \caption{ Difference in time-of-flight from the expected value for $K^-$ 
 (upper panel), $K^+$ (middle panel), and protons (lower panel), 
 for particles with momenta between 2.4 GeV and 2.5 GeV.  The black 
 curves are for all particles passing electron identification cuts, 
 while the blue curves have cuts on the SHMS aerogel, heavy gas detector 
 and calorimeter  detectors, as discussed in the text. The vertical 
 dashed lines indicate the timing cuts used for particle identification.
 }
 \label{fig:pkpid}
 \end{figure}

Kaons were identified by requiring a signal
less than 3 photoelectrons (p.e.) in the SHMS
aerogel detector~\cite{aero_ref} for momenta below the kaon
Cherenkov threshold of 2.9 GeV, with an 
efficiency of about 94\%. For positively
charged particles with momenta above 2.9 GeV, 
the aerogel signal was 
required to be above a threshold that increased
with momentum to give a constant efficiency
of about 92\%. This cut was used to reduce the flux 
of protons, which is much larger than the kaon flux. 
No cut on the aerogel signal was needed for negative 
particles with momenta above 2.9 GeV.
The signal size in the SHMS 
Heavy Gas Cherenkov (HGC) was required to be less
than 0.5 p.e. for all momenta, which greatly reduced
the pion flux above the HG pion threshold of 
2.65 GeV, with an efficiency of about 97\%. In order to 
further reduce pion contamination, particles 
that passed through the low-efficiency ``hole" in 
the middle of the HGC detector 
were not used (see Ref.~\cite{HallCSIDIS:2025plr}).
Finally, the kaon 
time-of-flight (TOF) was required to agree
within 0.5 nsec relative to that expected
for kaons traversing the 22 m flight path
between the target and the detectors in the
SHMS hut, with an efficiency of about 96\%, as illustrated
in Fig.~\ref{fig:pkpid}. It can be observed that the large 
tails from pions to the left of the kaon peaks are greatly
reduced by the kaon PID cuts (compare the blue and black 
curves). The relative energy deposited in the SHMS
calorimeter was required to be below 80\% of the 
deposition expected for electrons, with an efficiency
of $93\%$ to $97\%$, depending on momentum.  To keep pion and proton contributions to 
the kaon signals below 10\%, the kaon momentum was required to be below 3.6 GeV 
(3.2 GeV)  for positive (negative) kaons. The contamination from electrons or positrons 
was found to be negligible. 

The events generated by the Monte Carlo simulation (SIMC) were subjected to the same 
geometrical and fiducial cuts at for the real data. The simulation included a careful treatment of kaon decays passing through the spectrometer and detector hut. Energy loss and multiple scattering 
in the target and detector stack were taken into account. 
To evaluate radiative corrections, 
the cross section model used the
fragmentation functions of DSS~\cite{dss_ff} to simulate SIDIS contributions. 
Lacking a reliable model, the radiative tails from exclusive 
kaon electroproduction were not simulated. Based on 
preliminary information from the kaon form factor 
experiment (which also took data in 2018-2019)~\cite{KLTref}, we estimate that the 
relative exclusive tail contributions range from 0.5\% at low $(z,P_t)$ to 4\%
at high $(z,P_t$), based on our pion SIDIS studies.
No corrections were made for kaons originating from vector meson decays. 

We estimate a relative systematic uncertainty on kaon multiplicities of 4\%
for $K+$ and 6\% for $K^-$, dominated by the uncertainties in PID and 
radiative corrections. The errors on the azimuthal modulations are estimated to be
0.025, based on examining results with different PID cuts and radiative correction
models. 

\subsection{Proton identification and simulation}
Protons were identified by requiring an aerogel signal below 3.5 p.e., with an 
efficiency of 97\%, an HGC signal below 1 p.e. (98\% efficiency), and a TOF 
within 0.5 nsec of the expected time of arrival, with an efficiency of 97\%. 
The relative energy deposited in the SHMS
calorimeter was required to be below 80\%, with an efficiency
of about 99\%. The contamination of the proton signal from positrons, pions, and kaons 
was $<1\%$ at all momenta, due to the much longer travel time of protons compared to other particles traversing the spectrometer, as can be observed in Fig.~\ref{fig:pkpid}. 

The events generated by the Monte Carlo simulation (SIMC) were subjected to the same 
geometrical and fiducial cuts at for the real data. Energy loss and multiple scattering 
in the target and detector stack were taken into account.  The effects of hard
scattering reactions leading to a loss of events were not taken into account
explicitly, but are estimated to be less than 2\%.
To evaluate inelastic radiative corrections, we used
the empirical fit described below. 
The radiative tails from exclusive proton electroproduction were not simulated, nor were protons from vector meson decays. 

We estimate a relative systematic uncertainty on proton multiplicities of 4\%
dominated by the uncertainties in PID and 
radiative corrections. The errors on the azimuthal modulations are estimated to be
0.02, based on examining results with different PID cuts and radiative correction
models.

\subsection{Data Analysis}
 For each setting, the background- and accidentals-subtracted experimental yields were normalized by the accumulated beam charge. The normalized SIDIS electroproduction yield was corrected for all known inefficiencies of the two spectrometers such as the detector efficiencies, trigger efficiency, tracking efficiencies, computer and electronic live times.
 For each kinematic setting, the corrected yields were binned in $z$, the azimuthal angle ($\phi$), and transverse momentum ($P_t$). SIDIS multiplicities were determined by comparing the experimental and simulation yields, scaled by the multiplicity model used for the particular process. The simulation package took into account kaon decays, radiative effects, and position-dependent detector responses. For each bin in $(x,Q^2, z, P_t)$, the azimuthal distributions were fit with the 
 functional form
 \begin{equation}
M_0  [1 + A \cos(\phi^{*}) + B\cos(2\phi^{*})].
\label{eq:phidep}
\end{equation}

\section{Results for kaons}
\begin{figure}[hbt!]
\centering
\includegraphics[width=1.0\textwidth]{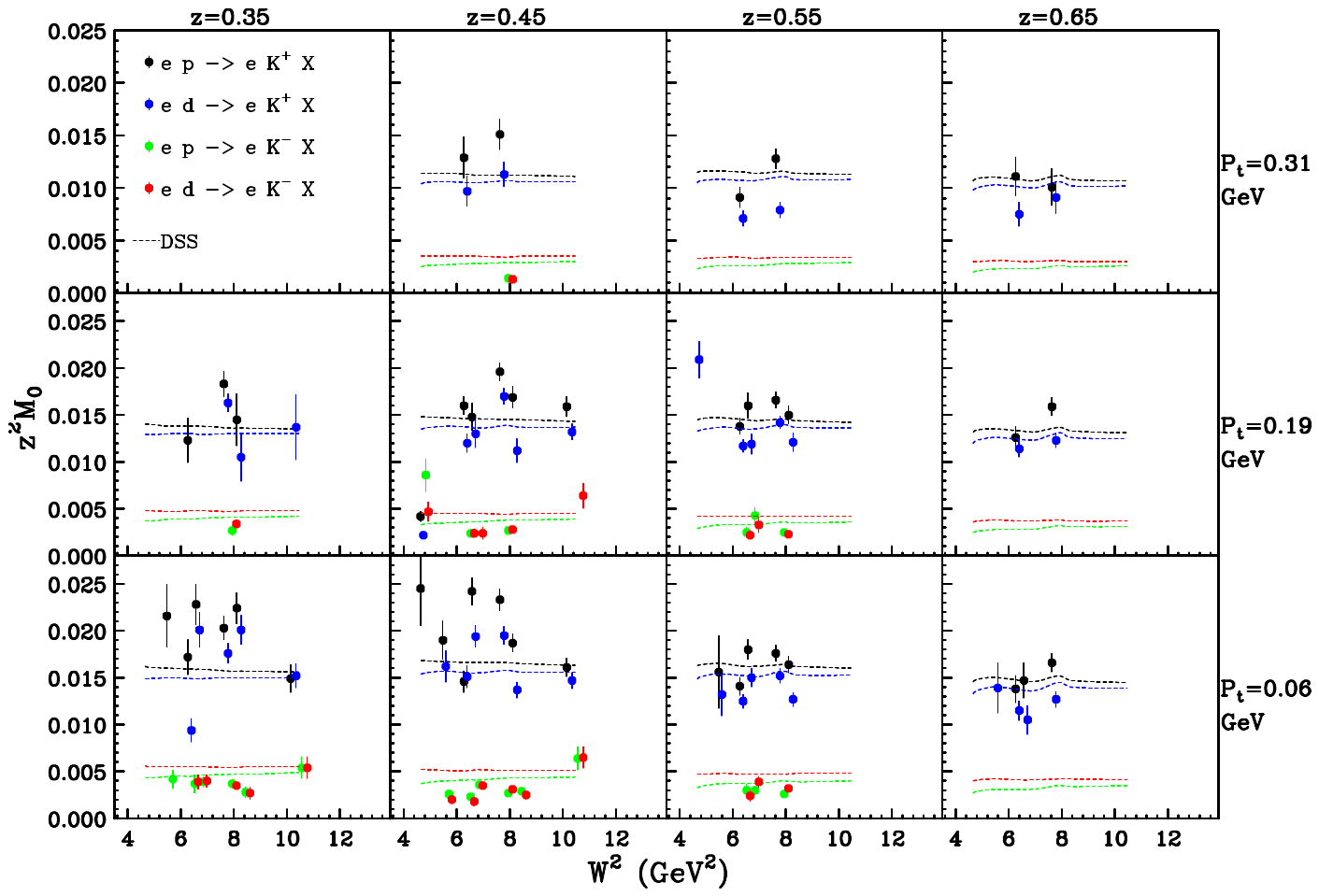}
 \caption{ Fit results for the parameter $M_0$ (see Eq. \ref{eq:phidep}) scaling the 
 kaon multiplicities as a function of $W^2$
 for four bins in $z$ (rows) and three bins in $P_t$ (columns).  
 The colors for the four reactions are indicated in the upper left panel. 
 The error bars include a 4\% (6\%)relative normalization systematic
 uncertainty for $K^+$ ($K^-$) added in quadrature with statistical uncertainties.
 The dashed curves
 are based on the DSS~\cite{dss_ff}
$P_t$-integrated kaon fragmentation functions, scaled by
the Gaussian $P_t$-dependence described in the text.
 }
 \label{fig:m0kw}
 \end{figure}

The values of $M_0$ scaled by $z^2$ are shown in 
Fig.~\ref{fig:m0kw} as a function of $W^2$ in a
four-by-three grid in $z$ and $P_t$. The results are
compared to predictions using CTEQ5~\cite{cteq5}
quark distributions and DSS~\cite{dss_ff} $P_t$-integrated
fragmentation functions, scaled by
the same exponential falloff as we determined from our
analysis of pion multiplicities~\cite{HallCSIDIS:2025plr}, 
given by $e^{-bP_t^2}$, with
$b=(0.185 + z^2 0.28)^{-1}$ GeV$^{-2}$.
The positive kaon multiplicities for the deuteron target are in remarkably
good agreement with CTEQ/DSS predictions, given that the kinematic settings
have a marginal affinity to TMD factorization (see Fig.~\ref{fig:rap}). The
$K^+$ multiplicities are
noticeably larger for the proton target than for the 
deuteron target, in contrast to the CTEQ/DSS predictions.
The $K^-$ multiplicities are much smaller
than the corresponding $K^+$ values, and also lie below the
predictions, except at the highest values of $W^2$, where the TMD affinity
is the highest.

\begin{figure}[hbt!]
\centering
\includegraphics[width=1.0\textwidth] {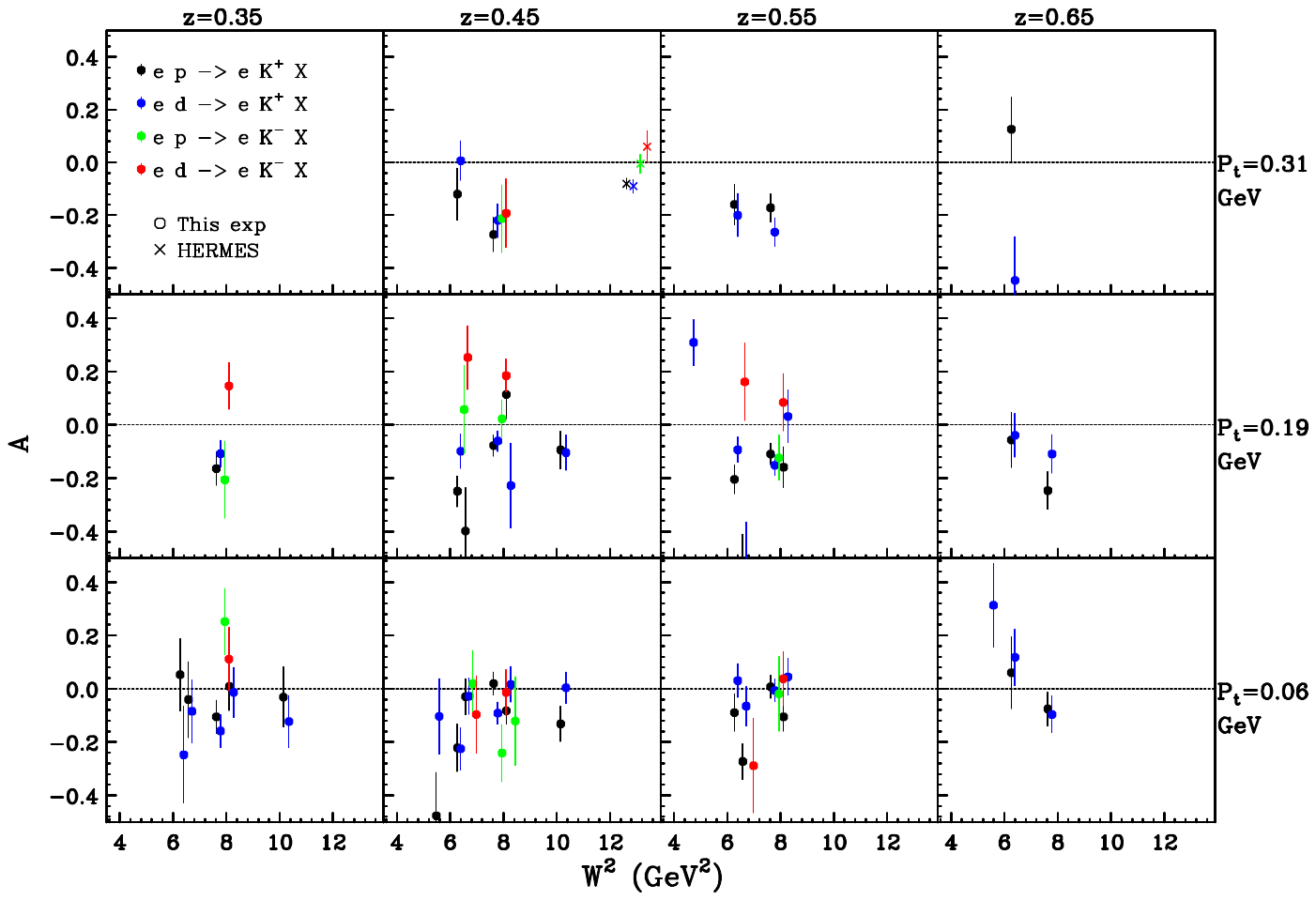}
 \caption{ Fit results for the parameter $A$ scaling the 
 kaon $\cos(\phi^*)$ dependence as a function of $W^2$
 for bins in $z$ and $P_t$. 
 The colors for the four reactions are indicated in
 the upper left panel. The solid circles are for this experiment,
 while the cross is from HERMES~\cite{HERMES:2012kpt}.
 The error bars include a 0.025 systematic
 uncertainty added in quadrature with statistical uncertainties.
  }
 \label{fig:phikw}
 \end{figure}

The results for  $A$, shown in 
Fig.~\ref{fig:phikw}, are small and consistent with zero, 
as was also found by HERMES~\cite{HERMES:2012kpt}. No dependence
on $W^2$ is apparent. We are
not aware of any theoretical evaluations with which to
compare our results. 

\begin{figure}[hbt!]
\centering
\includegraphics[width=1.0\textwidth] {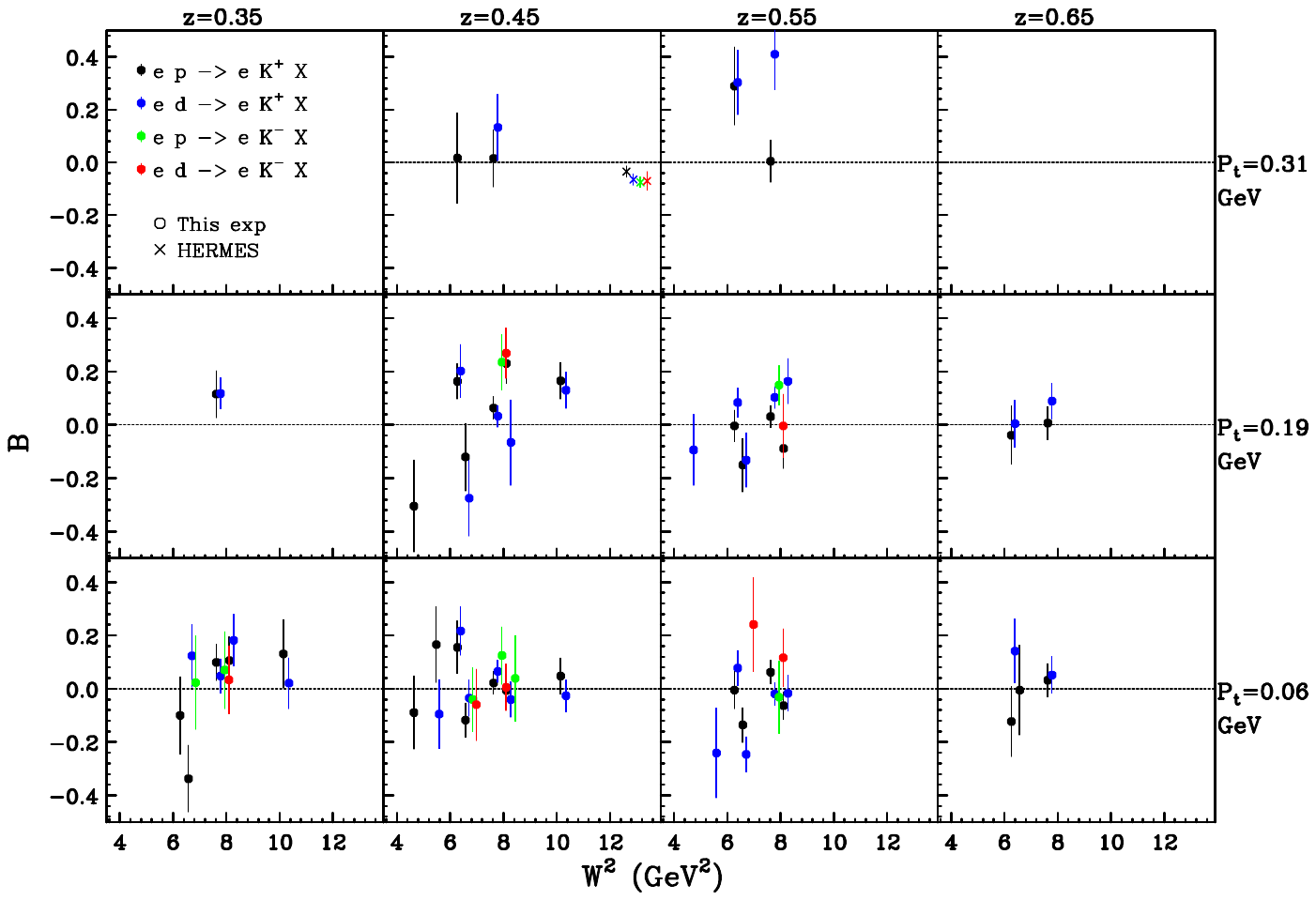}
 \caption{Same as Fig.~\ref{fig:phikw}, but in this case for
 the parameter $B$ describing the $\cos(2\phi^*)$ modulations.
 }
 \label{fig:2phikw}
 \end{figure}

The values of $B$ are shown in 
Fig.~\ref{fig:2phikw}. The values are close to zero, consistent
with results from HERMES~\cite{HERMES:2012kpt}. No
dependence on $W^2$ is apparent. We are not aware of any
theoretical predictions with which we can compare.

\begin{figure}[hbt!]
\centering
\includegraphics[width=1.0\textwidth] {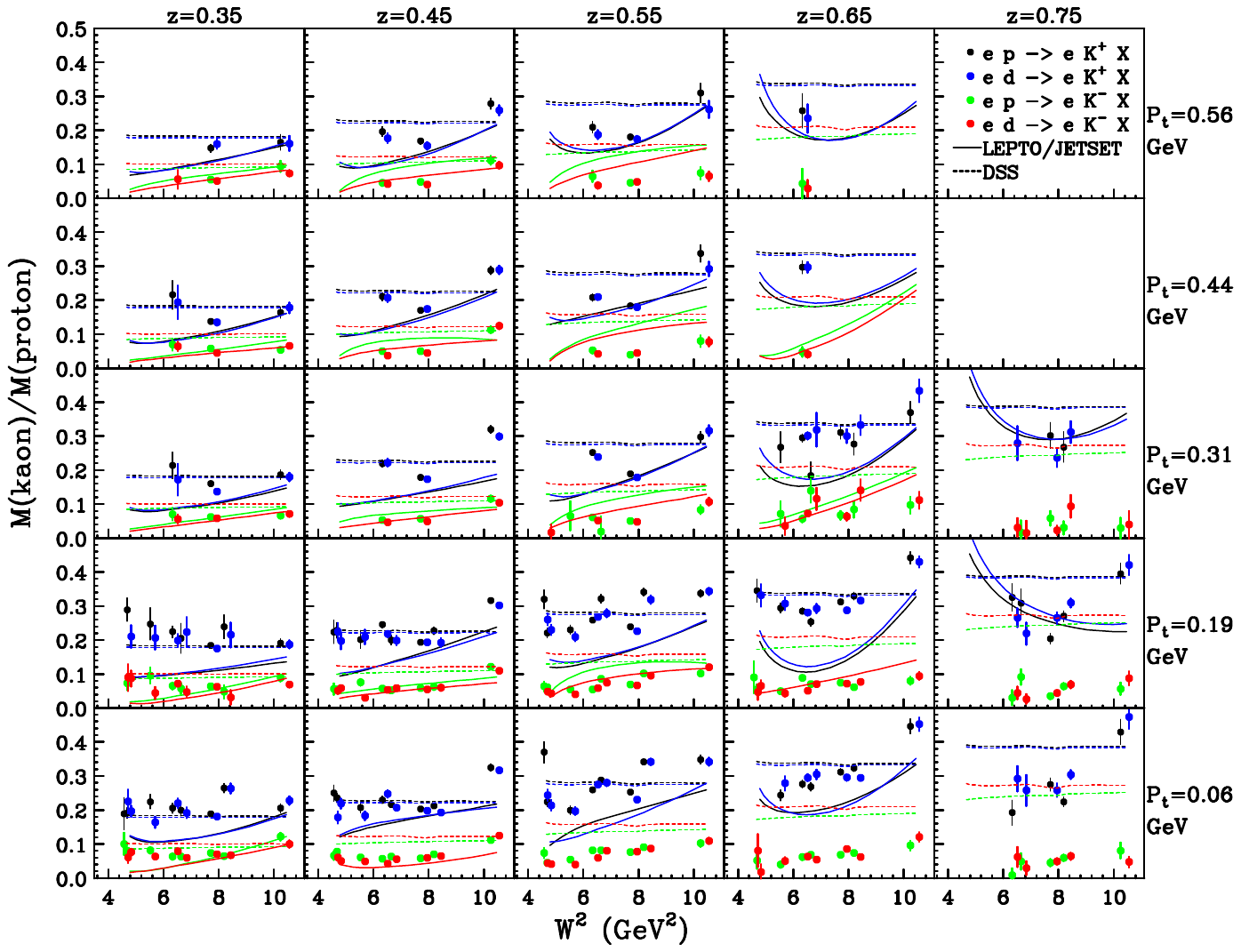}
 \caption{Ratios of $\phi^*$-averaged multiplicities for kaons compared 
 to pions~\cite{HallCSIDIS:2025plr,FF_paper} of the same sign charge as a function of $W^2$ in 
a grid in $z$ and $P_t$. The error bars are statistical only. 
The color codes for the
four reactions are given in the upper right panel.
The dashed curves are from the ratio of fragmentation functions from DSS~\cite{dss_ff}. The  solid curves are from the Lund Monte Carlo~\cite{Ingelman:1996mq,Sjostrand:1986hx}.
 }
 \label{fig:ratkw}
 \end{figure}

If we make the assumption that $A$ and $B$ are 
negligibly small, we can 
extend the coverage in $P_t$ to larger values, for which
only data near $\phi^*=180$ deg. were taken. 
In this case, we plot
in Fig.~\ref{fig:ratkw} the ratio of kaon 
to pion $\phi$-averaged yields, 
which allows us to 
compare directly with the Lund Monte Carlo 
simulation package LEPTO/JETSET~\cite{Ingelman:1996mq,Sjostrand:1986hx}, used
with input parameters from HERMES~\cite{HERMES:2012uyd}. 
For positively charged particles, the ratio 
increases gradually with
increasing $z$, roughly independent of $P_t$, 
and also exhibits a trend to increase with increasing $W^2$. 
These trends can also be observed in the Lund Monte Carlo, 
as well  as the CTEQ/DSS predictions. 
In contrast, the $K^-/\pi^-$ ratios are quite small at
all values of $z$, in increasing contrast with
both the Lund Monte Carlo and the CTEQ/DSS predictions.
Our ratios for $K^-/\pi^-$ are also smaller than 
those observed at  HERMES~\cite{HERMES:2012uyd}, 
albeit at higher $W^2$ than we reach
in the present experiment. However, a trend for our ratios 
to increase with increasing $W^2$ might provide for consistency.

\section{Results for protons}
\begin{figure}[hbt!]
\centering
\includegraphics[width=1.0\textwidth] {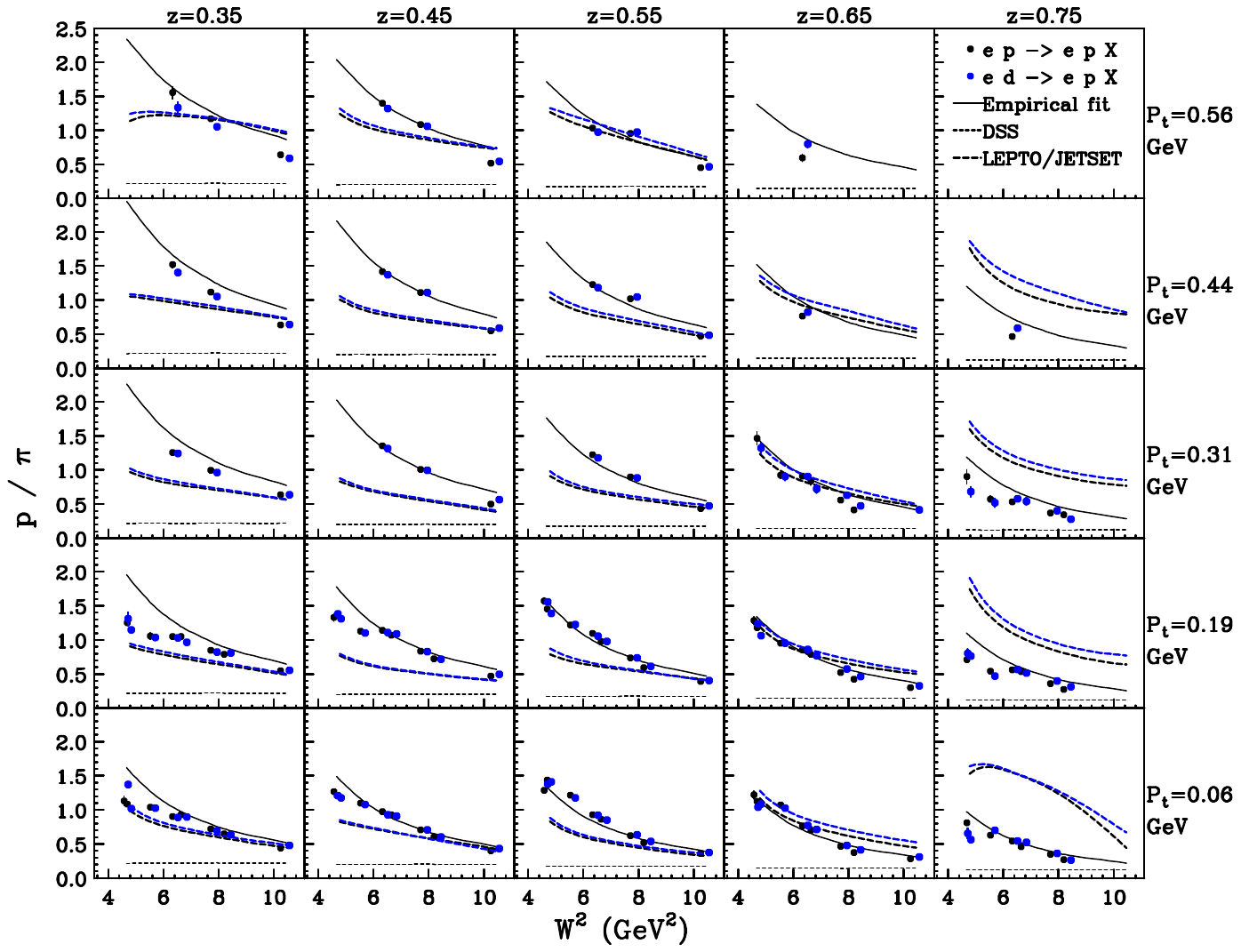}
 \caption{Ratios of $\phi^*$-averaged multiplicities for protons compared 
 to pions~\cite{HallCSIDIS:2025plr,FF_paper} as a function of $W^2$ in a
 grid in $z$ and $P_t$.  The black points are
 for $e p \rightarrow e p X$ and the blue points are for
 $e d \rightarrow e p X$. 
  The solid curves are an empirical fit, used
 in the evaluation of radiative corrections. The thin dashed curves are from the ratio of fragmentation functions from DSS~\cite{dss_ff}. The heavy dashed curves are from the 
 Lund  Monte Carlo~\cite{Ingelman:1996mq,Sjostrand:1986hx}.
 }
 \label{fig:ratpw}
 \end{figure}

The rates of protons were observed to be far higher than predicted
using DSS\cite{dss_ff} fragmentation functions, especially at
low values of $W^2$. In order to generate a new model to use in
the evaluation of radiative corrections, we first examined the
ratio of protons to pions. 
The ratios of $\phi^*$-averaged multiplicities for protons 
compared to pions~\cite{HallCSIDIS:2025plr,FF_paper} are shown in 
Fig.~\ref{fig:ratpw} as a function of $W^2$ in a
five-by-five grid in $z$ and $P_t$. In all cases,
a strong and smooth decrease with increasing $W^2$ can be
observed. No significant difference was observed between
the proton-to-pion multiplicity ratios between proton and deuteron targets. 
The solid curves are an empirical fit to the ratios, which
was used iteratively in the evaluation of radiative 
corrections. At the highest values of $W^2$, the ratios begin to
approach the ratios predicted from the ratio of DSS\cite{dss_ff} 
fragmentation functions,
consistent with an increasing affinity to a TMD description, as
discussed in the introduction. However, the observed ratios are still
at least a factor of two higher than the predictions at the highest
values of $W^2$. A much better description of
the observed ratios is provided by the phenomenological Lund Monte Carlo 
electroproduction model LEPTO/JETSET~\cite{Ingelman:1996mq,Sjostrand:1986hx}. While not
a perfect description, the strong decrease in the ratios with increasing
$W^2$ is evident.

\begin{figure}[hbt!]
\centering
\includegraphics[width=1.0\textwidth] {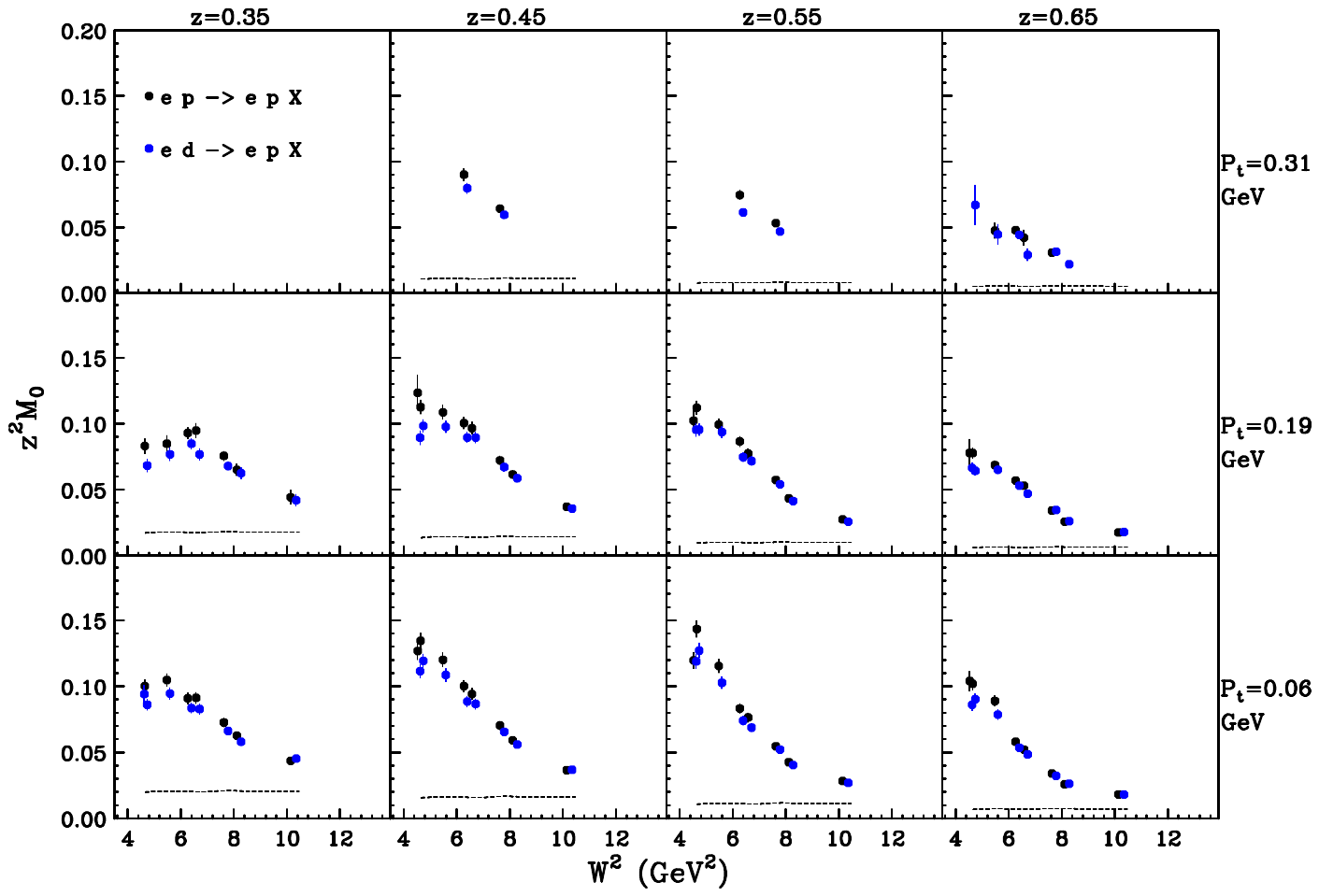}
 \caption{ Proton multiplicities (scaled by $z^2$) as a 
 function of $W^2$ in a
 grid in $z$ and $P_t$.  The black points are
 for $e p \rightarrow e p X$ and the blue points are for
 $e d \rightarrow e p X$. 
  The dashed curves are from the ratio of fragmentation functions from DSS~\cite{dss_ff}. The heavy dashed curves are from DSS~\cite{dss_ff}. }
 
 \label{fig:m0pw}
 \end{figure}

We have fit the data with a sufficient range in $\phi^*$
coverage to extract $M_0$, $A$ and $B$ for SIDIS protons.
The results for $M_0$, shown in Fig.~\ref{fig:m0pw}, again
show a strong decrease with increasing $W^2$, with little
dependence on target. This is in strong contrast to the
much lower predictions using CTEQ5~\cite{cteq5} parton 
distribution functions and
DSS\cite{dss_ff} fragmentation functions, which are
essentially independent of $W^2$ at fixed $z$ and $P_t$. 

\begin{figure}[hbt!]
\centering
\includegraphics[width=1.0\textwidth] {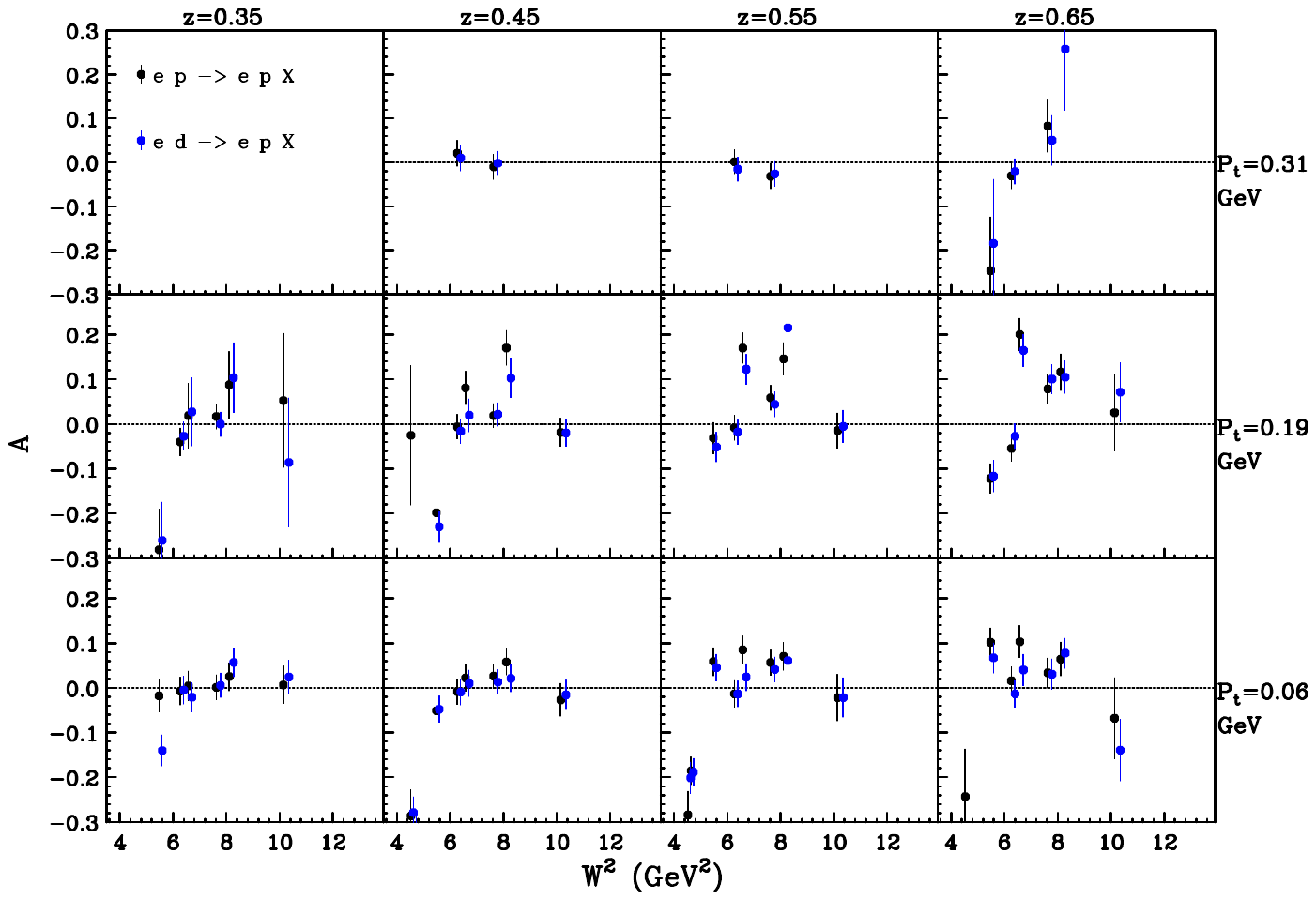}
 \caption{ Fit results for the parameter $A$ scaling the 
 proton $\cos(\phi^*)$ dependence as a function of $W^2$
 for bins in $z$ and $P_t$. The black points are
 for $e p \rightarrow e p X$ and the blue points are for
 $e d \rightarrow e p X$. The error bars include an estimated systematic
 error of 0.02 added in quadrature to the statistical errors.
 }
 \label{fig:phipw}
 \end{figure}

The values of $A$ are shown in 
Fig.~\ref{fig:phipw}. The results are close to zero for
$z<0.5$, and tend to be slightly positive at higher $z$. No
dependence on $W^2$ is apparent. We
have not been able to find any predictions to compare to.

\begin{figure}[hbt!]
\centering
\includegraphics[width=1.0\textwidth] {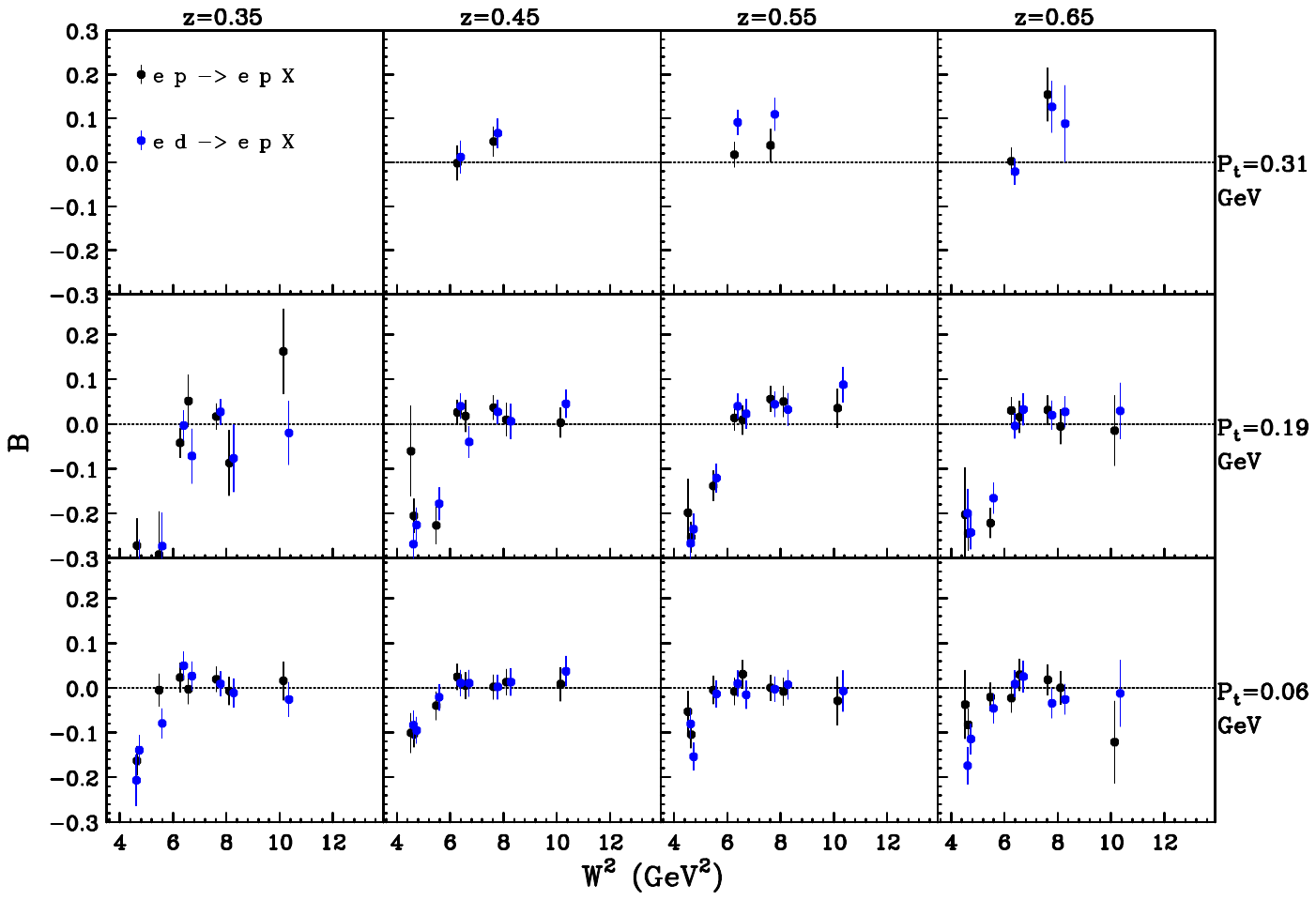}
 \caption{Same as Fig.~\ref{fig:phipw}, but in this case for
 the parameter $B$ describing the $\cos(2\phi^*)$ modulations.
  }
 \label{fig:2phipw}
 \end{figure}

The values of $B$ are shown in 
Fig.~\ref{fig:2phipw}. The values are very close to zero
at low $P_t$, and slightly positive at larger $P_t$. No
dependence on $W^2$ is apparent. We
have not been able to find any predictions to compare to.

\section{Data tables}
Numerical results for the multiplicities are tabulated in a full
 three-dimensional grid in $(z,P_t,\phi^*)$ for each
 target, and HMS setting in $(x,Q^2)$ on the
 Hall C SIDIS experimental results web page~\cite{webpage}. 
 Separate tables are provided for SIDIS kaons (29300 entries)
 and protons (24400 entries).
 In these tables, 
 each HMS spectrometer setting was divided in two, with relative
 scattering angle either positive or negative.  The table
 also includes results from eight additional HMS settings beyond
 those present in this paper, for which only the deuteron target
 was used. 
 The tables also include multiplicity results
 with no radiative corrections applied, which may prove useful
 in future global fits with consistent radiative correction models
 and formalism.


\section{Summary}

For SIDIS kaons, the kinematic range of this experiment spans the rapidity region between
the ``soft" central region and the region of good affinity for  a TMD factorization
description. For $K^+$, we find quite good agreement with predictions based on fragmentation
functions fitted to higher energy data. In contrast, we find many fewer $K^-$ than
predicted, especially at larger values of $z$, for all values of $W^2$. We find 
the azimuthal modulations for SIDIS kaons to be small and consistent with zero.

The kinematic range of the present experiment sits squarely 
in the ``soft" central rapidity region for SIDIS protons. This
provides an explanation of why the observed proton multiplicities
are factors of two (at the highest values of $W^2$) to ten
(at the lowest values of $W^2$) higher than predictions based
on fits to higher energy data. While the proton 
$\cos(2\phi^*)$ modulations are consistent with zero, there
is an indication of positive values of the $\cos(\phi^*)$
modulation. 

For both kaon and proton SIDIS, we find the HERMES-tuned
LEPTO/Jetset~\cite{Ingelman:1996mq,Sjostrand:1986hx}
event generator provides a qualitative 
description of the results, and with further tuning could
be a useful practical tool in understanding the non-perturbative
contributions to SIDIS.

\section{Acknowledgments}

We are grateful to M. Cerutti for providing calculations of the
2022 MAP model at each of the kinematic settings of this experiment.

This material is based upon work supported by the U.S. Department of Energy, Office of Science, Office of Nuclear Physics under Contract No. 89243126CSC000213.
This work was also funded in part by the U.S. Department of Energy, Office of Science, contract numbers DE-AC02-06CH11357, DE-FG02-07ER41528, DE-FG02-96ER41003, and by the U.S. National Science Foundation grants PHY 2309976, 2012430, 2013002 and 1714133 and the Natural Sciences and Engineering Research Council of Canada grant SAPIN-2021-00026. We wish to thank the staff of Jefferson Lab for their vital support throughout the experiment. We are also grateful to all granting agencies providing funding support to the authors throughout this project.

\bibliographystyle{apsrev4-1}
\bibliography{main}

\end{document}